\documentclass[preprint,10pt,numbers]{sigplanconf}

\usepackage{amsmath}
\usepackage{listings}
\usepackage{titlesec}
\usepackage[hyphens]{url}
\usepackage[hidelinks]{hyperref}
\usepackage{cleveref}
\hypersetup{breaklinks=true}

\begin{document}

\setlength{\pdfpageheight}{\paperheight}
\setlength{\pdfpagewidth}{\paperwidth}
\lstset{basicstyle=\ttfamily\footnotesize,breaklines=true}




\title{Dependently Typing R Vectors, Arrays, and Matrices}

\authorinfo{John Wrenn, Anjali Pal, Alexa VanHattum, Shriram Krishnamurthi}
           {Brown University}
           {shriram@brown.edu}

\maketitle

\begin{abstract}
  The R programming language is widely used in large-scale data
  analyses. It contains especially rich built-in support for dealing
  with vectors, arrays, and matrices. These operations feature
  prominently in the applications that form R's raison d'\^etre,
  making their behavior worth understanding. Furthermore, ostensibly
  for programmer convenience, their behavior in R is a notable
  extension over the corresponding operations in mathematics, thereby
  offering some challenges for specification and static verification.

  We report on progress towards statically typing this aspect of the R
  language. The interesting aspects of typing, in this case, warn
  programmers about violating bounds, so the types must necessarily be
  dependent. We explain the ways in which R extends standard
  mathematical behavior. We then show how R's behavior can be
  specified in LiquidHaskell, a dependently-typed extension to
  Haskell. In the general case, actually verifying library and client
  code is currently beyond LiquidHaskell's reach; therefore, this
  work provides challenges and opportunities both for typing R and for
  progress in dependently-typed programming languages.
\end{abstract}

\section{Retrofitting Types for R}

The R programming language~\cite{rlang} is widely used~\cite{rpop} in 
scientific computing applications. R is also a language with somewhat 
peculiar semantics~\cite{morandat} when compared with languages with a strong 
semantics tradition (such as typical functional languages). The unusual 
behaviors of R have proven to be a challenge for performance and are 
the subject of compiler and run-time systems research~\cite{kalibera}.

In this paper, we instead focus on static verification of R. Because R 
is a dynamic language, we are following the design principle labeled 
as \emph{retrofitting}~\cite{tejas-type-sys-js}: to turn as many 
dynamic errors as possible into static ones. This is the same principle 
followed by decades of research in dynamic languages, customized 
somewhat differently to each language. Specifically, this paper focuses 
on bounds checking.

Our bounds-checking work is embedded in a more general context. Our
type system has support for multiple R datatypes, and can therefore
also catch other standard errors such as the (incorrect) use of
non-function values in function application positions. Since
statically retrofitting such errors is well-known and we do not
provide any interesting insights into how these are different in R, we
do not discuss these further in the paper. Nevertheless, it is
important to note that the static information in this paper is
embedded in a conventional type system that also catches such errors.

The rest of this paper makes the following contributions:
\begin{enumerate}

\item As a primer for readers unfamiliar with R, we provide an
  overview of some of R's unusual behaviors. This identifies three kinds
  of information we want to record statically: \emph{shapes} (as expected)
  but also \emph{modes} and \emph{containers}.

\item We specify mode and container constraints and show how to embed them in an
  existing type system (Haskell's).

\item We specify shape constraints and show how to embed them in an
  existing type system (LiquidHaskell's). Together, these provide a
  new static semantics for these parts of R.

\item We present a brief description of a working toolchain that will
  transpile some R programs to use the above types.

\item We then discuss the difficulties these type specifications
  present to LiquidHaskell. For the most part, it is difficult to
  type either libraries or clients, making this a research challenge
  for dependent-typing language technology.

\item We also discuss ways in which our specifications can be made
  stronger.

\end{enumerate}

This paper represents the state of this work as of late
2016. Improvements to Liquid Haskell since that time will surely make
it possible to do this work at a much more ambitious level.

\section{A Primer on R Vectors, Arrays, and Matrices}
The surprising flexibility of R's vector operations leaves ample room 
for interpretations that, while seemingly robust, are confounded by 
simple counter examples. Consider the expression:
\begin{lstlisting}[language=R]
foo <- function(a,b) c(a, b)
\end{lstlisting}
Here, the name \lstinline{foo} is bound to a function value consuming two 
arguments. The function \lstinline{c} is a built-in function of R, used to 
combine values into a vector. For example, the expression 
\lstinline{c(1, 2, 3)} yields a vector containing three elements, 
\lstinline{1}, \lstinline{2}, and \lstinline{3}. 

Given the appearance of similar examples in online tutorials 
~\cite{rtut}, an R novice might assume that \lstinline{foo} is a 
function that consumes two arguments and produces a vector containing 
those elements. This interpretation can be reinforced with a few 
exploratory expressions in the R \textsc{REPL}. Calling \lstinline{foo} 
with the arguments \lstinline{1} and \lstinline{2} returns a vector 
containing those values:
\begin{lstlisting}[language=R]
> foo(3, 4)
[1] 3 4
\end{lstlisting}
Similarly, if we pass \lstinline{foo} two vectors, we can subscript its 
resultant twice:
\begin{lstlisting}[language=R]
> foo(c(3,4),c(5,6))[1][1]
[1] 3
\end{lstlisting}
However, these intuitions fail to explain that the application of 
\lstinline{foo} may, in fact, return a vector of any length, or that 
the result of the application of \lstinline{foo} may be indexed into 
endlessly (e.g., \lstinline{foo(c(3,4),c(5,6))[1][1][1]...}). We will 
therefore explain R's behavior in more detail in the rest of this 
section. This is a partial description of the behavior of a subset of R, and provides the basis for the more formal encoding of the behavior
discussed in later sections.

\subsection{Modes}
All values in R are tagged with an associated \emph{storage mode}. 
These resemble types in traditional languages, but because we will have 
a richer notion of types, we will use R's terminology and call these 
\emph{mode}s. The built-in function \lstinline{mode} can be used to 
inspect the storage mode of a value:
\begin{lstlisting}[language=R]
> mode(1)
[1] "numeric"
> mode(TRUE)
[1] "logical"
> mode("foo")
[1] "character"
> mode(function(a)a)
[1] "function"
\end{lstlisting}
In this paper, we consider a subset of R that includes the 
\lstinline{numeric}, \lstinline{logical}, \lstinline{character}, and 
\lstinline{function} modes, representing real numbers, booleans, 
strings, and functions, respectively. Additionally the 
\lstinline{numeric}, \lstinline{logical}, and \lstinline{character}
modes contain the constant \lstinline{NA}, used to represent missing 
data.

\subsection{Containers}

\paragraph{Vectors} 
What a programmer might think of as a primitive value is often actually 
a vector containing precisely one element. One-dimensional, homogeneous 
collections of data are R's lingua franca of data representation.
Instances of numeric, character, and logical literals in R source code 
can be understood as desugaring to vectors of length one containing 
those values; e.g., the numeric literal \lstinline{1} is a vector 
containing the component \lstinline{1}. R values remain boxed in vectors 
across operations, allowing for surprises such as subscripting 
a literal \emph{ad infinitum}:
\begin{lstlisting}[language=R]
> 5
[1] 5
> 5[1]
[1] 5
> 5[1][1][1][1][1]...
[1] 5
\end{lstlisting}

With vector-boxing in mind, the \textit{combination} function 
\lstinline{c} should not be thought of as a vector constructor, but as 
a vector \emph{concatenator}, as its arguments will themselves be 
vectors. For example, the aforementioned expression 
\lstinline{foo(c(1,2),c(3,4))} evaluates to a vector of length four, 
not two.

R's constant \lstinline{NULL} (not to be confused with \lstinline{NA}) 
is the empty vector:
\begin{lstlisting}[language=R]
> c() == NULL
[1] TRUE
> length(NULL)
[1] 0
\end{lstlisting}
In this work, we make the simplifying assumption that vectors have a
fixed size.

\paragraph{Arrays}
The \lstinline{array} type provides a facility for representing
\emph{n}-dimensional data in R. Array objects serve as a wrapper
around a vector containing the array's contents and a vector of
\emph{n} numeric values describing the dimensions of the
array. Accordingly, arrays are constructed with two vectors:
\begin{lstlisting}[language=R]
> array(c(1,2,3,4), c(2,2))
     [,1] [,2]
[1,]    1    3
[2,]    2    4
\end{lstlisting}
If the programmer does not supply enough elements to fill the specified 
shape, the elements are \emph{recycled} to cover the difference.
\begin{lstlisting}[language=R]
> array(c(1,2), c(3,3))
     [,1] [,2] [,3]
[1,]    1    2    1
[2,]    2    1    2
[3,]    1    2    1
\end{lstlisting}

\paragraph{Matrices}
The \lstinline{matrix} type is used for representing two-dimensional 
data and can be constructed using the \lstinline{matrix} function:
\begin{lstlisting}[language=R]
> matrix(c(1,2,3,4),c(1,4))
     [,1] [,2] [,3] [,4]
[1,]    1    2    3    4
\end{lstlisting}
Behaviorally, a matrix can be regarded as simply a name given to 
instances of array that are two-dimensional:
\begin{lstlisting}[language=R]
> class(matrix(c(1,2,3,4),c(1,2)))
[1] "matrix"
> class(array(c(1,2,3,4),c(1,2)))
[1] "matrix"
\end{lstlisting}

\subsection{Binary Operations}\label{primer-binops}
R's binary arithmetic operations (\lstinline{+}, \lstinline{-}, 
\lstinline{*}, \lstinline{/}, etc.) and binary logical operations 
(\lstinline{&}, \lstinline{|}, \lstinline{>}, \lstinline{==}, etc.) 
perform their respective operations \textit{pair-wise} on the elements 
of their arguments:
\begin{lstlisting}[language=R]
> 1 + 1
[1] 2
> c(1, 2, 3) + c(1, 2, 3)
[1] c(2, 4, 6)
\end{lstlisting}

\paragraph{Between Mixed Containers}
The left and right operands of a binary operation may be of different 
container types. The resultant of the operation is a vector if both 
operands are vectors, and an array if either operand is an 
array. 

\paragraph{Between Mixed Modes}
In binary logical operators, numeric arguments are coerced to logical 
values (with zero being falsey), and the resultant is always a 
collection of logical values.
\begin{lstlisting}[language=R]
> TRUE & 0
[1] FALSE
> TRUE & -2
[1] TRUE
\end{lstlisting}
In binary arithmetic operators, logical values are coerced to numeric 
values (\lstinline{TRUE} coerces to \lstinline{1} and \lstinline{FALSE} 
coerces to \lstinline{0}).
\begin{lstlisting}[language=R]
> 1 + FALSE
[1] 2
> 1 + TRUE
[1] 0
\end{lstlisting}
Character values cannot be coerced into logical or arithmetic modes.
\begin{lstlisting}[language=R]
> 1 + "R"
Error in 1 + "R" : non-numeric 
argument to binary operator
\end{lstlisting}

\paragraph{Between Mixed Sizes}
When the operands of a binary operation between containers are of 
different lengths, the elements of the shorter operand are recycled to 
cover the difference in lengths. The expression:
\begin{lstlisting}[language=R]
c(1, 2, 3, 4) * c(0, 1)
\end{lstlisting}
can be thought of as taking an intermediate step before which the 
pair-wise operation is performed:
\begin{lstlisting}[language=R]
> c(1, 2, 3, 4) * c(0, 1, 0, 1)
[1] 0 2 0 4
\end{lstlisting}
R issues a \emph{warning} if the longer operand is not a multiple of 
the shorter operand:
\begin{lstlisting}[language=R]
> c(1,2) + c(1,2,3)
[1] 2 4 4
Warning message:
In c(1, 2) + c(1, 2, 3) :
  longer object length is not a 
  multiple of shorter object length
\end{lstlisting}

\subsection{One-Dimensional Indexing}
\label{primer-indexing}
Both vectors and arrays support a one-dimensional indexing 
operation. Indexing a multi-dimensional array in a one-dimensional 
manner ignores the dimensioning rules of the array and indexes into the 
array's internal, one-dimensional vector of contents.
\begin{lstlisting}[language=R]
> array(c(1,2,3,4), c(2,2))[3]
[1] 3
\end{lstlisting}

\paragraph{With Positive Numerics}
When a container is indexed into with a container of positive 
numerics, R returns a vector containing the values at the indices 
corresponding to each of those positive numerics. This style of 
indexing can be used, for example, to produce a new vector from an 
input with that input's values in a different order:
\begin{lstlisting}[language=R]
> c("a", "b", "c")[c(3, 2, 1)]
[1] "c" "b" "a"
\end{lstlisting}

\paragraph{With Negative Numerics}
When a container is indexed into with a container of negative 
numerics, R returns a vector that \emph{excludes} all the values at 
those indices. 
\begin{lstlisting}[language=R]
> c("a", "b", "c")[c(-1, -3)]
[1] "b"
\end{lstlisting}

\paragraph{With Logical Values}
When a container is indexed into with a container of logicals, R 
applies the container of logicals as a mask over the vector, returning 
a container containing the elements beneath each \lstinline{TRUE} 
value. If the logical vector is shorter than the vector being indexed 
into, its values are recycled to fill the difference in lengths. A 
common use of this is masking every other element in a vector:
\begin{lstlisting}[language=R]
> c("a", "b", "c")[c(TRUE, FALSE)]
[1] "a" "c"
\end{lstlisting}

\subsection{Subscript-Assignment}
Containers may be mutated using the subscript assignment operator, 
which largely follows the behavior of single-dimension indexing with 
assignment. When the replacement vector is shorter than the the number 
of elements that are being replaced, the elements of the replacement 
vector are recycled, as demonstrated in the third \textsc{REPL} 
interaction:
\begin{lstlisting}[language=R]
> a <- c(1, 2, 3, 4)
[1] 1 2 3
> a[c(1,2)] = NA; a
[1] NA NA  3 4
> a[c(1,2,3,4)] = c(1, 2); a
[1] 1 2 1 2
\end{lstlisting}
Subscript-assignment operations, like the pair-wise binary operations,
give a warning when the right operand is not a multiple of the left 
operand:
\begin{lstlisting}[language=R]
> a[c(1,2)] = c(NA, NA, NA); a
Warning message:
In a[c(1, 2)] = c(NA, NA, NA) :
  number of items to replace is not 
  a multiple of replacement length
[1] NA NA  3
\end{lstlisting}
When the right operand is a \lstinline{NULL}, R gives an \emph{error}:
\begin{lstlisting}[language=R]
> a[1] = NULL
Error in a[1] = NULL : 
  replacement has length zero
\end{lstlisting}

\section{A Preliminary R Typing Workflow}

\paragraph{The Nature of Types}

These examples encompass a relatively small but rich subset of R for
which a formal encoding and automated verification tool would be
helpful. Moreover, they point towards a notion of types for R. The
restrictions on binary operations between containers seem expressible
as a relation between the operands' respective
\begin{itemize}
  \item container types,
  \item storage modes, and
  \item sizes.
\end{itemize}
These properties varyingly impact the type of the result of a binary 
operation. The container types of a binary operation's operands require 
that the binary operation is overloaded, as depending on the container 
types of its operands, the resultant may be a vector or an array. 
Whether the binary operation is an arithmetic or logical operation 
governs the storage mode of its result, and depending on the modes of 
its operands, it may be necessary to first coerce the mode of one or 
both of its operands the operation can be applied. Haskell's type 
system is sufficiently expressive to encode this interaction between 
types and behavior.

The sizes of a binary operation's operands, however, impose constraints 
on the sizes of the operands and indicate the size of its resultant. 
Encoding these constraints requires the expressiveness of a dependent 
type system. We use LiquidHaskell's refinement language as a 
specification language for stating constraints on sizes.

\paragraph{Target Language Choice}

Our choice of LiquidHaskell is driven by pragmatics. In principle, we
just need some dependently-typed language that can express these
constraints, and indeed many alternatives might have worked at least
as well. With LiquidHaskell, however, we get two benefits:
\begin{itemize}

\item We can exploit the expressiveness of the Haskell type system,
  which is already very powerful and offers the ability to encode
  several features of R; for instance, its type-classes enable a
  natural expression of R's overloading.

\item LiquidHaskell, which extends Haskell, is an increasingly mature
  language with \emph{full checker automation}. Automation is vital to
  a workflow that starts in another language. If instead we had used
  an interactive proof assistant, then R programmers would be asked to
  write proof steps for a significantly mangled program (e.g., to
  encode state monadically) in a language that does
  not at all resemble R.

  In particular, LiquidHaskell's automation is built atop SMT~\cite{jhala}
  (satisfiability modulo theories). The theories provide the checker
  an extensible framework. If we find certain patterns in R that do
  not naturally type, we can in principle add theories to the solver
  that will discharge these patterns. In that sense, we can view
  LiquidHaskell's refinements as a convenient interface to an SMT
  solver to solve R's array constraints, rather than having to build
  that translation ourselves.

\end{itemize}
Of course, these arguments are not canonical: they do not rule out the
use of other dependently-typed target languages. In addition, there
are clearly non-trivial user interface issues raised by this
translation, and making the errors decipherable by an R programmer
remains wide open. Nevertheless, Haskell and LiquidHaskell combined
offer a fine grounds for \emph{prototyping} a solution, which might
then perhaps be translated into a native solution for R itself.

\paragraph{The Pipeline}

These specifications have enabled us to begin construction of LiquidR, a
preliminary pipeline for typing a subset of R programs. Our system
parses R programs, reduces them to a core language, and then compiles
that reduction to Haskell. The compiled result utilizes an alternative
prelude that implements many of the datatypes and operations that are
part of our subset of R. Given the specifications described in the
subsequent sections, we can in principle perform two kinds of checks:
\begin{enumerate}

\item Most importantly, we can check that an R program does not
  violate these type constraints by compiling it and applying the
  Haskell and LiquidHaskell type checkers to the result. This assumes
  that our specifications are sufficiently faithful to R's semantics
  (which we try to assure by exploring numerous examples).

\item In addition, we can check that our re-implementations of R's
  vector, array, and matrix operations in Haskell faithfully implement
  these specifications. These would, of course, only be of interest to
  Haskell programmers who want access to R's semantics in Haskell.

\end{enumerate}
Note that the former checks the \emph{client} (the R program using R's
built-in operations), while the latter checks the \emph{server} (a
Haskell implementation of these operations). Each of these has
independent value, and of course a LiquidHaskell user using the
Haskell implementation of R's operations would use the specifications
for both purposes.

Due to the limitations of LiquidHaskell we have been able to make
only preliminary progress on both fronts, as we discuss further in
\cref{pipeline-dev}.

\section{Encoding R in (Liquid)Haskell}
The type of each particular mode is encoded using Haskell's 
\lstinline{Maybe} type, which has the variants \lstinline{Just} (used to
represent the presence of a value), and \lstinline{Nothing} (used to
represent the absence of a value). We use \lstinline{Nothing} to
encode R's \lstinline{NA}:
\begin{lstlisting}[language=Haskell]
type Logical   = Maybe Bool
type Numeric   = Maybe Double
type Character = Maybe Double
\end{lstlisting}
To characterize these types as being modes, we create and instantiate
a typeclass \lstinline{Mode}:
\begin{lstlisting}[language=Haskell]
class Mode a
instance Mode Logical
instance Mode Numeric
\end{lstlisting}
Each operation between single values of a given mode is specified as 
instance of a typeclass describing that operation:
\begin{lstlisting}[language=Haskell]
class (Mode a, Mode b, Mode c) 
    => Addition a b c | a b -> c  
  where mode_add :: a -> b -> c
\end{lstlisting}
For example, adding two numerics either yields an integer or, if either 
argument is a \lstinline{Nothing}, an \lstinline{NA}:
\begin{lstlisting}[language=Haskell]
instance Addition Numeric 
                  Numeric
                  Numeric where
  mode_add (Just l) (Just r) = Just (l + r)
  mode_add _ _ = Nothing
\end{lstlisting}

\subsection{Container Representation}
\paragraph{Vectors}
Vectors are encoded as Haskell lists:
\begin{lstlisting}[language=Haskell]
type Vector a = [a]
\end{lstlisting}

\paragraph{Arrays}
We represent arrays as a pair of lists, the first storing the
dimensions of the array, and the second storing the members of the
array:
\begin{lstlisting}[language=Haskell]
type Array  a = ([Int],[a])
\end{lstlisting}
However, this Haskell type fails to preclude the storage of 
conceptually flawed arrays; there may be, for example, too few or too 
many elements in relation to the capacity that the dimensions suggest, 
or the array may be instantiated with negative dimensions. We enforce 
that the number of elements is equal to the product of the dimensions 
using a LiquidHaskell refinement:
\begin{lstlisting}[language=Haskell]
type Array a = ([Nat],[a])<{
    \shape elems ->
      (product shape) = (len elems)
  }>
\end{lstlisting}

\paragraph{Matrices}
Matrices, being simply a name given to arrays of two dimensions, are not
distinguishable from arrays at a Haskell type level, since Haskell is
not sufficiently expressive enough to constrain the fist component of
the \lstinline{Array} tuple to lists of two elements. Therefore, as a 
Haskell type, \lstinline{Matrix} is an alias for \lstinline{Array}:
\begin{lstlisting}[language=Haskell]
type Matrix a = Array a
\end{lstlisting}
We encode the additional dimensioning restriction by extending our
LiquidHaskell refinement for \lstinline{Array}:
\begin{lstlisting}[language=Haskell]
type Matrix a = ([Nat],[a])<{
    \shape elems ->
      (product shape) = (len elems) &&
      (len shape) = 2
  }>
\end{lstlisting}

\paragraph{Abstract Representation}
Since vectors, arrays, and matrices can be used interchangeably in
many contexts, we characterize container-ness using the
Haskell typeclass \lstinline{Container} and provide instances of it for both
\lstinline{Vector} and \lstinline{Array}:
\begin{lstlisting}[language=Haskell]
class Container m where
  length :: m -> Int

instance Container (Vector a) where
  length = ...
  
instance Container (Array a) where
  length = ...
\end{lstlisting}

\subsection{Container Construction}
\paragraph{Vectors} 
Although we use Haskell's lists to encode R vectors, it is insufficient
to compile instances of the R \lstinline{c} function to Haskell list
literals, since the \lstinline{c} function \emph{concatenates} its
arguments. The length of \lstinline{c}'s output is the sum of the
lengths of its inputs. We encode this notion of length by creating the
LiquidHaskell measure \lstinline{llen}, which consumes a list of lists
and produces the sum of their lengths:
\begin{lstlisting}[language=Haskell]
measure llen :: [[a]] -> Int
  llen ([])   = 0
  llen (x:xs) = (len x) + (llen xs)
\end{lstlisting}
We compile instances of R's \lstinline{c} function to the Haskell
function \lstinline{combine}. We use the \lstinline{llen} measure to
provide a refinement on the length of the result of \lstinline{combine}:
\begin{lstlisting}[language=Haskell]
combine :: 
 ls:([[_]]) -> 
 {l:([_]) | (len l) = (llen ls)}
\end{lstlisting}

\paragraph{Arrays}
Arrays in R are constructed using
the \lstinline{array} constructor function (rather than literal
syntax). This consumes a vector of 
dimensions and a vector of elements and produces an instance of the 
\lstinline{array} class. Here, too, it is not suitable to compile 
occurrences of the \lstinline{array} constructor directly to our Haskell 
representation of arrays because using our \lstinline{Array} data-type would 
make for an \emph{over}-constrained constructor. Recall that if an array 
constructor is supplied with enough elements to fill the specified 
shape, the elements are \emph{recycled} to cover the difference:
\begin{lstlisting}[language=R]
> array(c(1,2), c(3,3))
     [,1] [,2] [,3]
[1,]    1    2    1
[2,]    2    1    2
[3,]    1    2    1
\end{lstlisting}
Accordingly, we transpile occurrences of the \lstinline{array} 
constructor to applications of a function called
\lstinline{array}, which only constrains that the array must be
non-empty (via this LiquidHaskell refinement):
\begin{lstlisting}[language=Haskell]
array :: (Mode t) =>
    shape:(NonEmptyContainer NonNegNumeric) ->
    elems:(NonEmptyContainer t) ->
    reslt:(Array t)
\end{lstlisting}
This constraint allows for this R runtime error to be statically
reported:
\begin{lstlisting}[language=R]
> array(c(),c())
Error in array(c(), c()) : 
  'data' must be of a vector type, was 'NULL'
\end{lstlisting}

\paragraph{Matrices}
It is not an error to specify more than two dimensions to the R 
\lstinline{matrix} constructor; however, the resulting array will be
truncated to the first two given dimensions. If we believe that such
truncations are likely the result of programmer error, we can report
so statically by supplying a LiquidHaskell refinement for our Haskell
\lstinline{matrix} constructor that rejects superfluous dimensions:
\begin{lstlisting}[language=Haskell]
matrix :: (Mode t) =>
    shape:(ContainerNElems NonNegNumeric 2) ->
    elems:(NonEmptyContainer t) ->
    reslt:(Matrix t)
\end{lstlisting}

\subsection{Binary Operations}

Recall from \cref{primer-binops} three types of constraints 
(mode, container type, and size) are imposed on 
the arguments of operations on containers. To encode constraints on the 
operands' modes and container types, we construct a Haskell typeclass 
for each each binary operation. In this section, we'll consider an 
encoding of addition:
\begin{lstlisting}[language=Haskell]
class (Container a, 
       Container b, 
       Container c) 
      => Addition a b c | a b -> c  
  where add :: a -> b -> c
\end{lstlisting}
All valid combinations of containers and mode for a given operation are 
encoded as instances of that operation's typeclass. For example, we 
identify addition between a vector of numeric values and an array of 
logical values as an acceptable operation by instantiating the Addition 
typeclass with those types:
\begin{lstlisting}[language=Haskell]
instance Addition (Vector Mode.Numeric) 
                  (Array  Mode.Logical) 
                  (Array  Mode.Numeric) 
  where add = ...
\end{lstlisting}
Note that the parametrization of this instance for \lstinline{Addition}
denotes that the expected result of an addition between a vector and an
array is an array.

Although the binary operations do not impose size constraints on the 
operands, LiquidHaskell's ability to bless clients using operations 
that \emph{do} impose size constraints is predicated on how strongly it 
can reason about the sizes of data produced by operations. Accordingly, 
we introduce a LiquidHaskell refinement to describe binary operations 
that leave the operands unconstrained and refine the length of the 
result. In the LiquidHaskell refinement below, \lstinline{Arithmetic} 
is a type describing a function that, when it consumes two non-empty 
containers, produces a container whose length is equal to that of the 
longest operand. If either operand is empty, the result will be empty:
\begin{lstlisting}[language=Haskell]
type Arithmetic t u v =
  (Container t, Container u, Container v) =>
    a:t -> 
    b:u ->
   {c:v | if (len a) = 0 || (len b) = 0
          then (len c) = 0
          else if (len a) >= (len b)
            then (len c) = (len a)
            else (len c) = (len b)}
\end{lstlisting}
The size of a binary operation's operands does not produce any hard 
errors, but recall that a \emph{warning} is produced if the length of 
the larger operand is not a multiple of the length of the shortest 
operand. We can strengthen our refinement to statically report this 
warning as a type error by using the modulus of the operands' lengths to 
test multiplicity:
\begin{lstlisting}[language=Haskell]
type Arithmetic t u v =
  (Container t, Container u, Container v) =>
    a:t -> 
    b:u ->
   {c:v | if (len a) = 0 || (len b) = 0
          then (len c) = 0
          else if (len a) >= (len b)
               && (len a) mod (len b) = 0
            then (len c) = (len a) 
          else if (len a) < (len b)
               && (len b) mod (len a) = 0
            then (len c) = (len b)
          else false}
\end{lstlisting}

\subsection{Subscript Operations}
Other container operations, such as subscripting, are also encoded 
using distinct Haskell typeclasses, e.g.:
\begin{lstlisting}[language=Haskell]
class (Container a, 
       Container b)
      => Subscript a b
  where subscript :: a -> b -> a
\end{lstlisting}
and the acceptable operations are encoded as instances:
\begin{lstlisting}[language=Haskell]
instance (Container a) 
    => Subscript a (Vector Mode.Numeric)
  where subscript = ...
instance (Container a) 
    => Subscript a (Vector Mode.Logical)
  where subscript = ...
\end{lstlisting}
Since we do not yet consider arrays with named-dimensions, we omit an 
instance parametrized with \lstinline{(Vector Mode.Character)}. Like 
the binary operations, subscript operations do not report errors (or 
even warnings). Nonetheless, we refine their result type with the 
expected size so LiquidHaskell can reason about those values 
elsewhere in the program.

\paragraph{With Positive Numerics}
Recall from \cref{primer-indexing} that when a vector or array 
is indexed with a container of non-negative numerics, R returns a 
container that holds the values at the indices corresponding to each of 
those positive numerics. The result is therefore a container of the 
same length as the subscript:
\begin{lstlisting}[language=Haskell]
type SubscriptNonNegative t = 
  (Container t) =>
    a:t ->
    b:(Container NonNegNumeric) ->
   {c:t | (len c) = (len b)}
\end{lstlisting}

\paragraph{With Negative Numerics}
Recall that when subscripting with a container of negative numerics, 
R returns a vector that excludes all of those values at those 
indices. R considers the subscript vector as a set (so excluding an
element more than once is equivalent to excluding it once), and ignores
exclusions outside the bounds of the array. The length of the result is
not related to the length of the subscript, but rather the number of
distinct elements indices within the bounds of the array in the
subscript. This can be encoded in a LiquidHaskell refinement, but it
increases the burden on the constraint solver. It is useful, then, to
instead express the weaker constraint that a negative numeric
subscript operation cannot produce an output longer than the vector
being subscripted. 
\begin{lstlisting}[language=Haskell]
type SubscriptNonPositive t = 
  (Container t) =>
    a:t ->
    b:(Container NonPosNumeric) ->
   {c:t | (len c) <= (len b)}
\end{lstlisting}

\paragraph{With Logicals}
Finally, recall that when subscripting with a container of logicals, R 
applies the container of logicals as a mask over the vector, returning 
a container that holds the elements beneath each \lstinline{TRUE} 
value. If the number of \lstinline{TRUE} or \lstinline{NA} values in the
subscript is greater than the length of the vector being subscripted,
subscripting with logicals can produce a result longer than the
container being subscripted on:
\begin{lstlisting}[language=R]
> c(1,2,3)[c(TRUE,TRUE,TRUE,TRUE)]
[1]  1  2  3 NA
\end{lstlisting}
If the number of \lstinline{TRUE} or \lstinline{NA} values in the
subscript is less than the number of values in the container being
subscripted, this operation can produce a result shorter than the
container being subscripted on:
\begin{lstlisting}[language=R]
> c(1,2,3)[c(FALSE,TRUE,FALSE,FALSE)]
[1]  2
\end{lstlisting}
Without stating properties about the contents of the subscript, it is impossible
to create a meaningful refinement of the length of this operation's result.

\paragraph{Haskell Integration}
Encoding indexing presents an integration challenge. If we overload the 
indexing operation using Haskell type classes, it is not possible to 
define two instances of \lstinline{Subscript} for a \lstinline{Numeric} 
subscript such that those instances can then be independently refined 
with \lstinline{SubscriptNonNegative} and 
\lstinline{SubscriptNonPositive}; Haskell will reject these typeclass 
instances as overlapping. Since only a single instance for 
\lstinline{Subscript} can be defined with \lstinline{Numeric} in the 
subscript position, and that instance may only be refined with a single 
LiquidHaskell type, we refine this instance by combining the 
\lstinline{SubscriptNonNegative} and \lstinline{SubscriptNonPositive} 
refinements into a single refinement:
\begin{lstlisting}[language=Haskell]
type SubscriptNumeric t =
  (Container t) =>
    a:t ->
    b:{bv:(Container Numeric) | 
        (allNonNegative b) 
     or (allNonPositive b)} ->
   {c:t | if (allNonNegative b) 
          then (len c) == (len b)
          else if (allNonPositive b) 
          then (len c) <= (len b)
          else false}
\end{lstlisting}

\subsection{Subscript-Assignment}
As with previous operations, we encode acceptable mode and container
combinations of the subscript-assignment operation with a Haskell
type class and appropriate instances.
\begin{lstlisting}[language=Haskell]
class (Container a, 
      Container b, 
      Container c, 
      Container d)
   => SubscriptAssignment a b c d | a b c -> d
 where subscriptAssign :: a -> b -> c -> d
\end{lstlisting}
Although subscript-assignment largely borrows the behavior of subscript,
it introduces both a warning and an error related to the size of the
replacing value. We will consider each variant in turn:

\paragraph{With Postive Numerics}
This is expressive enough to report the multiplicity warning and NULL 
assignment error statically:
\begin{lstlisting}[language=Haskell]
type SubscriptAssignmentNonNeg t v =
 (Container t, Container v) =>
   a:t                         -> 
   b:(Container NonNegNumeric) ->
  {c:v | (len c) > 0
         if (len a) >= (len b)
           then ((len a) mod (len b)) = 0
           else ((len b) mod (len a)) = 0} ->
  {d:t | if (len b) > (len a)
           then (len d) = (len b)
           else (len d) = (len a)}
\end{lstlisting}
However, it is accurate only with our assumption that vectors are of a
fixed size. The length of the resultant is, in fact, equal to the
highest numeric value in \lstinline{b}.

\paragraph{With Negative Numerics}
Does not report multiplicity warning; doing so requires knowing the set
difference between the indices of \lstinline{a} and the absolute values 
of the members of \lstinline{b}.
\begin{lstlisting}[language=Haskell]
type SubscriptAssignmentNonPos t v =
 (Container t, Container v) =>
   a:t                         ->
   b:(Container NonPosNumeric) ->
  {c:v | (len c) > 0 }         ->
  {d:t | (len a) == (len b)}
\end{lstlisting}

\paragraph{With Logicals}
Does not report multiplicity warning; doing so requires knowing how many
elements in \lstinline{a} were masked by \lstinline{b}.
\begin{lstlisting}[language=Haskell]
type SubscriptAssignmentLogical t v =
 (Container t, Container v) =>
   a:t                         ->
   b:(Container Logical)       ->
   c:(Container NonPosNumeric) ->
  {d:t | (len a) == (len b)}
\end{lstlisting}

\section{From Here to TheRe}
\subsection{Pipeline Development}\label{pipeline-dev}
At the time of writing, we have been able to use our R-to-Haskell 
pipeline to cross-compile modest R programs within this subset. For variations 
on the specifications presented in this paper, we have additionally had 
success using this pipeline to expose basic runtime errors in R 
statically. An R program such as
\begin{lstlisting}[language=R]
foo <- function(a,b) c(a,b)
a <- foo(1,c(NA,3))
b <- foo(c(1,NA),3)
a + b
\end{lstlisting}
which evaluates to a vector containing \lstinline{2}, \lstinline{NA}, 
\lstinline{6} is transpiled into the Haskell program:
\begin{lstlisting}[language=R]
import LiquidR
program = do
  let foo = \\(x,y) -> combine [x,y]
  let a = foo ([Just 1.0],
               combine [[Nothing], 
                        [Just 3.0]])
  let b = foo (combine [[Just 1.0], 
                        [Nothing]],
               [Just 3.0])
  a `add` b
  endOfProgram
\end{lstlisting}

\paragraph{Haskell Type Checking} Using our compilation pipeline, the 
Haskell encoding of R's vector and array types presented in this paper 
is sufficient for reporting basic mode mismatches and function arity 
mistmatches statically. The Haskell typechecker, run on our pipeline's 
compilation of the following programs, is able to report a Haskell 
typechecking error for each of R's dynamic errors:
\begin{lstlisting}[language=Haskell]
> foo(1)
Error in a(1, 2, 3, 4): 
  unused arguments (1, 2, 3)
> foo(1,2,3)
Error in foo(1, 2, 3):
  unused argument (3)
> "bar"(1,2)
Error: 
  attempt to apply non-function
> "cat" + 1
Error in "cat" + 1 : non-numeric argument to binary operator
\end{lstlisting}

\paragraph{LiquidHaskell Type Checking} 
We are in the progress of integrating our LiquidHaskell specifications 
into this pipeline. Unfortunately, while our Haskell type encoding 
relies heavily on typeclasses, LiquidHaskell support for typeclass 
refinements, a recent addition to LiquidHaskell, remains preliminary. 
We have identified a handful of bugs that prevent us from 
annotating binary operations on typeclasses and are in the process of 
reporting these to the LiquidHaskell team.

We have integrated a modification of these refinements with a minimal 
standard library that does not support operator overloading (e.g., the 
only containers are vectors and the only mode is numeric), and used the 
LiquidHaskell typechecker to statically report basic 
size-related errors.

\subsection{Broadening the Scope}

We envision three levels of broadening the scope of this work, the
first of which should be straightforward, the second of which is
significant work but appears feasible, and a third of which is nearly
impossible.
\begin{enumerate}

\item The most natural extension is to cover even more of the
  constraints on vectors, arrays, and matrices. We have discussed some
  of these limitations. The main reason we have not explored them yet
  is because of the difficulty of getting LiquidHaskell to
  type-check, so we lack confidence in the correctness of richer
  assertions. Once the bugs in that system are fixed and we can check
  these assertions, we do not see major obstacles in even more
  precisely capturing R's behavior.

\item The next natural step is to cover more of R's datatypes. Some, like
  functions, we have only partially covered; though we can catch some
  function errors, there are many features of R's functions---such as
  optional arguments, named arguments, \emph{abbreviated} named
  arguments ~\cite{oddities}---that we have not covered. In addition, there are
  several more features of R, such as data frames (a multi-dimensional
  association of names with data), or R's \emph{three} different
  object systems. Naturally, modeling each of these will require
  significant elbow grease.

\item Most subtly, however, there is the problem of modeling the core
  language itself. In describing a compiler, we have implicitly
  assumed that R has a fixed language that can be subject to a static
  compilation system. This assumption is true of most languages, but
  it is not true of R. 

  In R, even presumed keywords and built-in constructs, like \lstinline{if},
  function definition, and assignment, are actually \emph{functions},
  and an initial library binds them to specific definitions. Because
  they are just bound variables, their binding can be changed by
  mutation, and they are globally re-defined.

  We dub this behavior \emph{antisyntactic}, since the language
  effectively goes out of its way to thwart syntax-driven static
  processing. As a result, our compiler \emph{assumes} that language
  constructs have their default meaning and have not been
  redefined. Creating a static type system that can correctly model
  the true behavior of R in the presence of syntax changes seems a
  fairly overwhelming task.

\end{enumerate}

In the immediate future, we plan on improving our 
compilation pipeline and Haskell encoding of R's types to model the 
mutating effects of subscript-assignment operations. The ubiquity of 
vectors and vector operations in R is half of the language's recipe for 
idiomatic code; the other half is the association of names with values. 
We would like to explore transpiling R programs using functions with 
named parameters. The dimensions of vectors and arrays, too, may be 
named; though at the cost of significantly increasing the complexity of 
our refinements, we suspect that LiquidHaskell has sufficient 
expressive power for this. Solutions for these problems may point to an 
effective compilation strategy for data frames.

\section{Related Work}

While there has been significant research on scripting languages in
general, some of which has inspired us, most of it does not directly
apply here. Similarly, literature on run-time system performance
improvements in R is typically expressed at a fairly low level of
abstraction, significantly below the specifications in this paper.

Morandat et al.~\cite{morandat} describe a formal semantics for a
large but only partially overlapping subset of R. Their work does not
delve into the minutiae of R's vector and array operations. Their
formal semantics does not model arrays, or logical and character moded
values, and only considers single-dimension non-negative numeric
indexing where the subscript is a vector of length one. We therefore
view our paper as a complement to theirs, providing a partial
semantics to a different corner of the R language. In addition, much
of our attention has gone towards encoding the semantics in a solver
with an eye towards a type-checking pipeline, which their work does
not cover.

Slepak, et al.'s~\cite{slepak} typed array-based language Remora is a
clean-slate design of an array programming language. Remora therefore
offers an elegant core language that could be a good translation
target for a subset of R. However, Remora does not have mutation,
limiting its applicability, and has limitations on
indexing. Furthermore, it does not currently offer strong solver
support. Therefore, while it is a promising platform, we chose to not
use it as our compilation target.

\section{Concusion}

This paper represents a preliminary approach to typing R
statically. Based on the behavior of critical R constructs like
vectors, arrays, and matrices, we have created a set of dependently
typed specifications that capture much of that behavior. We have also
created a preliminary pipeline to consume R source programs, compile
them into LiquidHaskell, and automatically run the type-checker.

Unfortunately, at this point our process works only for very modest
programs. Our major difficulty has been with LiquidHaskell: since our
translation needs to employ brand new features in the language, we
have hit untested and erroneous corners. As a result we do not yet
know whether, once these bugs are fixed, these assertions can be
checked quickly---though some of our others have not---or whether this
task will expose new challenges for dependent type solvers (which may
require some augmentation to the underlying decidable theories). 

At any rate, we believe this experiment is useful on several fronts:
for typing R, for exercising LiquidHaskell (by presenting it with
problems that are clearly dissimilar enough from the ones imagined by
the authors to trigger errors), and possibly also for putting solvers
through their paces. In particular, progress on the latter two fronts
to type R will have positive impacts on the many other tasks that also
depend on these systems.

We also believe this project will have value for other languages. Some
other scientific data processing languages also have unusual semantics
that extend that of conventional mathematics; those too would benefit
from our specifications. If they have a more tractable semantics, that is not
a problem: it simply makes the specification simpler (for instance, we
can simply delete parts of our specification) and might also make
solving the resulting constraints quicker. Therefore, it would be
worth comparing our specifications against those for libraries like
SciPy~\cite{scipy}, which might also profitably be compiled by a toolchain
similar to ours.

Finally, the use of a rich type system opens up other possibilities as
well. For instance, though R does not have a native ``unit'' system,
clearly units and dimensional analysis are vital checks in scientific
computations. It would therefore be interesting to consider extensions
to our specifications that enable units to be encoded and computed
along with other refinements of the values.


\acks

We thank Alex St.~Laurent for clarifying the relationship to the
Slepak, et al.~semantics, and Justin Pombrio for proofreading and
feedback.

\newpage

\bibliographystyle{abbrvnat}
\bibliography{cites}

\begin{thebibliography}{10}
\providecommand{\natexlab}[1]{#1}
\providecommand{\url}[1]{\texttt{#1}}
\expandafter\ifx\csname urlstyle\endcsname\relax
  \providecommand{\doi}[1]{doi: #1}\else
  \providecommand{\doi}{doi: \begingroup \urlstyle{rm}\Url}\fi

\bibitem[rla(2016)]{rlang}
The {R} project for statistical computing, 2016.
\newblock URL \url{https://www.r-project.org/}.

\bibitem[rpo(2016)]{rpop}
R passes {SAS} in scholarly use, June 2016.
\newblock URL
  \url{http://r4stats.com/2016/06/08/r-passes-sas-in-scholarly-use-finally/}.

\bibitem[rtu(2016)]{rtut}
{R} tutorial: Combining vectors, 2016.
\newblock URL \url{http://www.r-tutor.com/r-introduction/vector}.

\bibitem[Jhala(2014)]{jhala}
R.~Jhala.
\newblock Refinement types for {Haskell}.
\newblock In \emph{Proceedings of the ACM SIGPLAN 2014 Workshop on Programming
  Languages Meets Program Verification}, PLPV '14, pages 27--27, New York, NY,
  USA, 2014. ACM.
\newblock ISBN 978-1-4503-2567-7.
\newblock \doi{10.1145/2541568.2541569}.
\newblock URL \url{http://doi.acm.org/10.1145/2541568.2541569}.

\bibitem[Jones et~al.(2001--)Jones, Oliphant, Peterson, et~al.]{scipy}
E.~Jones, T.~Oliphant, P.~Peterson, et~al.
\newblock {SciPy}: Open source scientific tools for {Python}, 2001--.
\newblock URL \url{http://www.scipy.org/}.

\bibitem[Kalibera et~al.(2014)Kalibera, Maj, Morandat, and Vitek]{kalibera}
T.~Kalibera, P.~Maj, F.~Morandat, and J.~Vitek.
\newblock A fast abstract syntax tree interpreter for {R}.
\newblock In \emph{Proceedings of the 10th ACM SIGPLAN/SIGOPS International
  Conference on Virtual Execution Environments}, VEE '14, pages 89--102, New
  York, NY, USA, 2014. ACM.
\newblock ISBN 978-1-4503-2764-0.
\newblock \doi{10.1145/2576195.2576205}.
\newblock URL \url{http://doi.acm.org/10.1145/2576195.2576205}.

\bibitem[Lerner et~al.(2013)Lerner, Politz, Guha, and
  Krishnamurthi]{tejas-type-sys-js}
B.~S. Lerner, J.~G. Politz, A.~Guha, and S.~Krishnamurthi.
\newblock {TeJaS}: Retrofitting type systems for {JavaScript}.
\newblock In \emph{Dynamic Languages Symposium}, 2013.

\bibitem[Morandat et~al.(2012)Morandat, Hill, Osvald, and Vitek]{morandat}
F.~Morandat, B.~Hill, L.~Osvald, and J.~Vitek.
\newblock Evaluating the design of the {R} language: Objects and functions for
  data analysis.
\newblock In \emph{Proceedings of the 26th European Conference on
  Object-Oriented Programming}, ECOOP'12, pages 104--131, Berlin, Heidelberg,
  2012. Springer-Verlag.
\newblock ISBN 978-3-642-31056-0.
\newblock \doi{10.1007/978-3-642-31057-7_6}.

\bibitem[Mortel(2006)]{oddities}
T.~Mortel.
\newblock Wondrous oddities: {R}'s function-call semantics, 1 2006.
\newblock URL
  \url{http://blog.moertel.com/posts/2006-01-20-wondrous-oddities-rs-function-call-semantics.html}.

\bibitem[Slepak et~al.(2014)Slepak, Shivers, and Manolios]{slepak}
J.~Slepak, O.~Shivers, and P.~Manolios.
\newblock \emph{An Array-Oriented Language with Static Rank Polymorphism},
  pages 27--46.
\newblock Springer Berlin Heidelberg, Berlin, Heidelberg, 2014.
\newblock ISBN 978-3-642-54833-8.
\newblock \doi{10.1007/978-3-642-54833-8_3}.
\newblock URL \url{http://dx.doi.org/10.1007/978-3-642-54833-8_3}.

\end{thebibliography}


\end{document}